\documentstyle[pre,aps,draft,psfig]{revtex}
\begin{document}

\tighten
\twocolumn[\hsize\textwidth\columnwidth\hsize\csname
@twocolumnfalse\endcsname

\title{Mini-stop bands in single-defect photonic crystal waveguides}
\author{Mario Agio$^{1,2}$ and Costas M.\ Soukoulis$^{1,3}$} 
\address{$^1$Ames Laboratory and Department of Physics and Astronomy,\\
Iowa State University, Ames, IA 50011}
\address{$^2$INFM-Dipartimento di Fisica ``A.\ Volta'', Universit\`a di Pavia, 
I-27100, Pavia, Italy}
\address{$^3$Research Center of Crete, P.O. Box 1527, 71110  Heraklion, 
Crete, Greece}

\date{\today }
\maketitle
\begin{abstract}
We numerically study single-defect photonic crystal waveguides obtained 
from a triangular lattice of air holes in a dielectric background.
It is found that, for medium-high air filling ratios, the transmission has 
very small values in narrow frequency regions lying inside the photonic 
band gap - the so-called mini-stop bands. Two types of mini-stop bands are 
shown to exist; one of which is due to the multimode nature of the waveguide.
Their dependence on the length of the waveguide and on the air filling ratio 
is presented. 
\end{abstract}
\pacs{42.70.Qs,42.82.Et}
]

Photonic band gap (PBG) materials, also known as photonic crystals (PCs), 
have been extensively studied recently because of their ability to control 
the propagation of light\cite{joan_book,souk_book,joan_nat}. 
A very promising application of PBGs is for improving the performance of 
waveguides\cite{mekis_prl,lin_science,baba_elett,tokushima_apl,loncar_apl,smith_apl,labill_prl,chow_nat}. 
A photonic crystal waveguide is basically a PC with a linear defect, which 
allows the propagation of light in a specific direction. 
PC waveguides provide a superior guiding mechanism  with respect to 
dielectric or metallic waveguides since they are ideally loss-less, 
because of the PBG properties.
Guiding the light without losses, and even through sharp corners 
using two-dimensional (2D) PCs was studied theoretically\cite{mekis_prl} 
and experimentally, both in microwaves\cite{lin_science} and in 
optical\cite{baba_elett,tokushima_apl,loncar_apl,smith_apl} regimes. 
To eliminate out-of-plane losses in 2D PC structures, one has to use a 
index-guiding mechanism in the vertical 
direction\cite{labill_prl,chow_nat,chutinan_prb,johnson_prb}.

Detailed numerical studies of straight PC waveguides of different widths,  
in a high dielectric background\cite{benisty_jap,olivier_pcic} 
and in air\cite{mekis_prb}, have revealed that a complete frequency gap, 
called mini-stop band (MSB) or mode-gap, exists for the 
guided modes of the waveguide. The term MSB\cite{olivier_pcic} refers to 
an anti-crossing mechanism between guided modes of different order; 
whereas the term mode-gap\cite{mekis_prb} is used for the band splitting 
of the folded guided mode at the edge of the one-dimensional (1D) 
Brillouin zone. Indeed, a straight PC waveguide is actually a system with 
a discrete periodicity along the waveguide's axis, in which the 1D 
periodic potential given by the PC may produce mode couplings whenever 
the Bragg condition is fulfilled. 
Existence of MSBs has also been experimentally demonstrated 
by Olvier {\em et al.}\cite{olivier_pcic} for waveguides in GaAs based PC 
heterostructures. In all of the previous studies no MSBs were observed 
for the 2D PC waveguide, which has one row of cylinders missing 
(single-defect waveguide). We adopt the term MSB 
to refer both the anti-crossing and the band splitting of a folded mode.

The purpose of this work is to investigate the single-defect PC waveguide 
for a triangular lattice of air holes in a dielectric background, 
showing the occurrence of MSBs.
Their dependence on the length of the waveguide and on the air 
filling ratio will be presented and connections with the Fabry-Perot 
modes of the 1D cavity made of the linear defect 
will be discussed. The dielectric 
constant $\epsilon$ of the background is chosen to be 11.56, which 
corresponds to GaAs at the wavelength $\lambda$ of $1.55  \mu m$. 
The 2D PC has a gap for the TE mode, whose width depends\cite{joan_book} 
on the air filling ratio $f$. TE is the mode which has the magnetic field 
parallel to the axes of the air cylinders. 
One can easily create a waveguide by removing a row of air holes. 
This structure is shown in Fig. 1, where $a$ is the lattice constant 
and $w$ is the width  of the waveguide. 
The waveguide has variable length in the dense $\Gamma-K$ direction 
($\hat{\bf x}$ direction). The field confinement in the perpendicular 
$\hat{\bf y}$ direction ($\Gamma-M$) is provided by 5 PC rows arranged 
at both sides of the waveguide. Numerical results have been obtained with 
a 2D finite-difference time-domain (FDTD) method\cite{taflove_book}, 
implemented with Liao absorbing boundary conditions\cite{liao_ss}.
The numbers 1 and 2 in Fig. 1 represent the positions of the FDTD 
source: position 1(2) means transmission along the $\hat{\bf x}
(\hat{\bf y})$ direction.\\ 
\begin{figure}[h]
\centerline{\psfig{figure=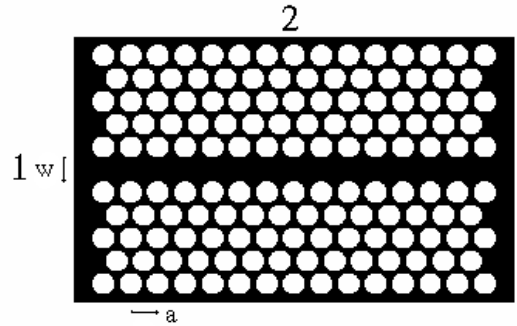,width=8cm}}
\caption{Triangular lattice of air holes in GaAs 
($\epsilon=11.56$). The missing holes yield a waveguide of width 
$w$ along the $\Gamma-K$ direction; $a$ is the lattice constant; 
$1$ and $2$ mark the positions of the source.}
\end{figure}
In Fig. 2, we present the results for the 
case of $f=60\%$ for both the perfect periodic case, as well as 
the waveguide case (source in position 1). Inside 
the gap of the periodic photonic crystal, the transmission coefficient 
of the waveguide is above 80\%\cite{comment1}, for almost all frequencies. 
Notice that there are two frequencies $\nu_{1}=0.321, \nu_{2}=0.377$ 
(units of $c/a$), where the transmission is low.
We attribute these gaps in the 
transmission to MSBs, i.e. gaps in the dispersion relation. 
Notice also, that, as we increase the length of the waveguide, both 
the drops of the MSBs get larger, as expected. On the other hand,
for frequencies corresponding to a guided mode, the transmission is 
independent of the length of the waveguide\cite{comment1}. 
Since the size of the PBG and the width $w$ are related to $f$, 
we have addressed the spectral behaviour of the transmission for 
different values of $f$.\\ 
\begin{figure}[h]
\centerline{\psfig{figure=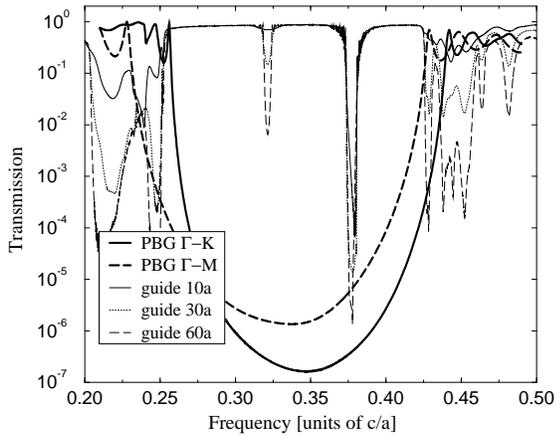,width=8cm,angle=270}}
\caption{Transmission spectra for various lengths of the waveguide, 
with $f=60\%$. The bold solid (dotted) line delimits the $\Gamma-K$ 
($\Gamma-M$) PBG of the bulk PC. The solid, dotted and dashed 
lines correspond to the transmission along the waveguide (source 
in position 1) with length $10a$, $30a$ and $60a$ respectively.}
\end{figure}
In Fig. 3, we plot the width of the 
gap for the periodic PC, as well as the frequencies $\nu_{1}$ and 
$\nu_{2}$ of the two MSBs versus $f$. 
From Fig. 3, one sees that for small $f$, there is no MSB 
inside the gap of the periodic system. This is the reason that no MSBs 
were seen in previous studies\cite{benisty_jap,olivier_pcic,mekis_prb} 
of single-defect PC  waveguides. Notice that for large $f$, more 
than two MSBs are seen. These additional MSBs present a dispersion 
with $f$ similar to the MSBs found for $f=60\%$.

To understand the character of the MSBs, we plot in Fig. 4  
the intensity of the electric field $|E^{2}|$ at the mid-gap frequency 
for the low- and high-frequency MSB for the case $f=60\%$. 
In Fig. 4(a), the profile extends all over the waveguide and it is 
a ``Bloch-like'' wave, similar to the pattern of the fundamental 
guided mode. However, it slowly decays as the field propagates through 
the waveguide. 
Notice that the PC waveguide is a 1D periodic system with a unit cell 
of length $a$, whose reduced Brillouin zone is $(0,\pi/a)$.
The modulated dielectric constant of the PC, in addition to the PBG 
confinement, entails also a periodic potential, which shall couple 
guided modes satisfying the Bragg condition.
\begin{figure}[h]
\centerline{\psfig{figure=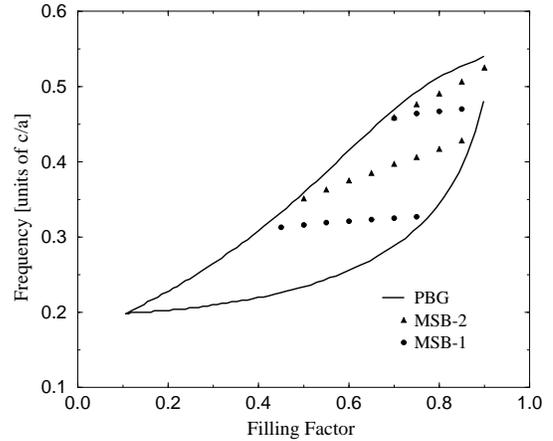,width=8cm,angle=270}}
\caption{Position of the MSBs versus $f$. MSB-1 (MSB-2) is for 
the type low-frequency (high-frequency) MSB of Fig. 2. 
The solid lines give the width of the H gap for different $f$.}
\end{figure}
\begin{figure}[h]
\centerline{\psfig{figure=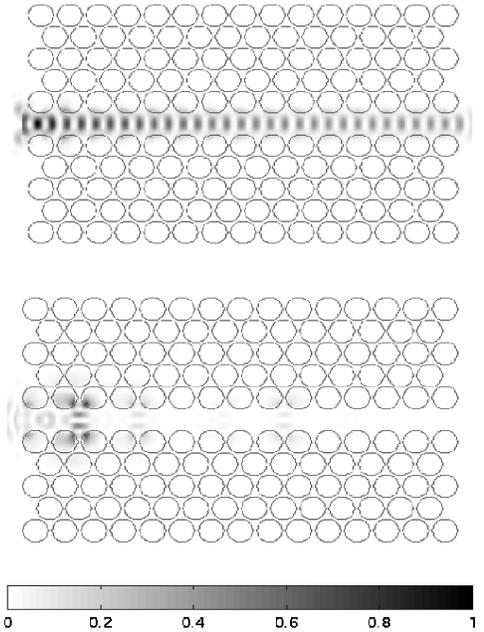,width=8cm}}
\caption{Normalized intensity of the electric field for the frequencies 
corresponding to the (top panel) low-frequency MSB and (bottom panel) 
high-frequency MSB of Fig. 2.}
\end{figure}
Returning to Fig. 4(a), if one counts the nodes in the field pattern 
along $\hat{\bf x}$, it is found that the wavelength is roughly equal 
to $a$; i.e. the unfolded wave-vector $k$ is $2\pi/a$, corresponding 
to $k=0$ in the first Brillouin zone. At $k=0$, the periodic 
potential couples the two counterpropagating fundamental modes 
($k=\pm 2\pi/a$), which would be otherwise degenerate.
Therefore, with Fig. 2 and Fig. 4(a) in mind, we recognize this MSB as 
the gap in the folded fundamental guided mode for $k=0$. 
\begin{figure}[h]
\centerline{\psfig{figure=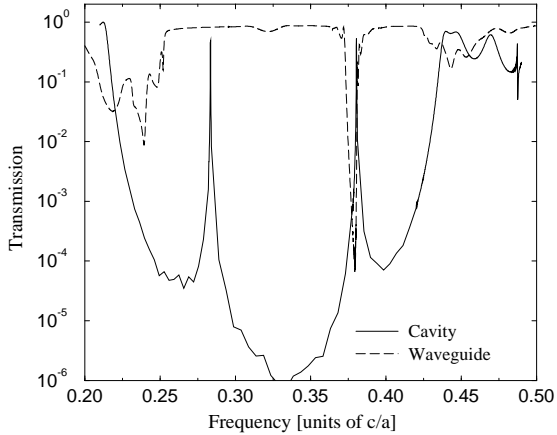,width=8cm,angle=270}}
\caption{Transmission spectra for the 1D cavity (solid line) 
and the waveguide (dotted line), with $f=60\%$ and length $10a$. 
For the 1D cavity is intended that the transmission is calculated 
along the $\bf \hat{y}$ direction, with the source in 
position 2, see Fig. 1.}
\end{figure}
The profile shown in Fig. 4(b) is completely different than that shown 
in Fig. 4(a). The electric field is now strongly localized and has nodes 
along $\hat{\bf y}$. Above the frequency of the MSB, the field pattern 
returns to be that of the fundamental guided mode. 
This suggests the existence of another gap in the dispersion relation 
of the fundamental mode. According to Fig. 4, the origin of the 
high-frequency MSB cannot have the same nature of the low-frequency MSB, 
otherwise the field patterns would be similar. We suspect that this MSB 
is due to a coupling (anti-crossing) between the fundamental mode and a 
higher mode. To check this idea, we have to find evidence of 
at least one higher guided mode at frequencies inside the PBG. 
Since there is a correspondence between the cut-off 
frequency of a guided mode and the resonant frequency of a 
Fabry-Perot mode\cite{olivier_pcic,kliewer_pr}, 
the search for cavity modes indirectly proves the existence of guided modes. 

The PC waveguide of Fig. 1 can be seen as a 1D cavity when the source 
is in position 2, see Fig. 1, and the progagation is along the 
$\hat{\bf y}$ direction. We have calculated the transmission 
coefficient for such 1D cavity and show the results in Fig. 5. 
As one can clearly see, there are two peaks in the transmission, which 
correspond to two cavity modes. The high frequency peak is very close 
to the MSB; the low-frequency peak does not align with any MSB. 
Looking at the field patterns of the cavity modes, we indeed find 
that the high-frequency peak has the same symmetry and the same number 
of nodes as the field for the MSB at $\nu_{2}$.
We infer that the MSB stems from the anti-crossing between the fundamental 
mode and the higher mode with a cut-off frequency given by the 
corresponding cavity mode.  
The low-frequency cavity mode is odd with respect to the 
$\hat{\bf x}$ axis; for the PC waveguide of Fig. 1 the odd modes cannot 
couple to even modes\cite{benisty_jap,olivier_pcic}. 
This is the reason for the absence of a MSB close to the ``odd'' 
resonant frequency. 
\begin{figure}[h]
\centerline{\psfig{figure=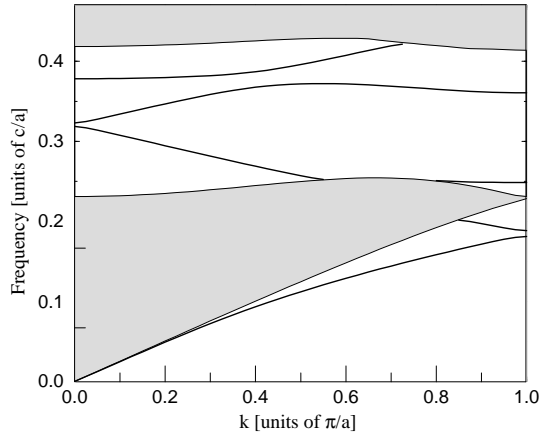,width=8cm,angle=270}}
\caption{Dispersion relation for the PC waveguide, with $f=60\%$. 
The gray area is the projected band structure of the 
perfect 2D crystal. The bold solid lines correspond to TE guided modes, 
even with respect to the waveguide's axis.}
\end{figure}
To give more insight of the MSBs, we have also calculated the band structure 
for the even TE modes of the waveguide shown in Fig. 1, with $f=60\%$. 
The results are plotted in Fig. 6. The horizontal axis is the wavector in 
the direction of the guide, and the band structure is presented in the 
reduced Brillouin zone scheme. The gray areas are the projections of
every mode in the band structure of the perfect crystal.
We focus only on the guided modes ranging in the PBG 
frequencies. The low-frequency MSB
is coming from the folding of the fundamental mode at 
$k$=0, as deducted from the previous field analysis. 
The high frequency MSB is coming from the anti-crossing between the 
fundamental and the higher order mode. 
This guided mode has a cut-off frequency $\nu_{c}=0.378$ and 
a small negative dispersion.\\
The weak depedence of the low-frequency MSB on f, shown in Fig. 3,
relies on the fact that the fundamental mode is almost concentrated 
in the dielectric channel. Therefore, its dispersion relation feels 
the filling factor only through the width $w$; likewise 
is for the position of the low-frequency MSB. On the other hand, the higher 
modes are more extended in the PC region and their dispersion relation 
will be more sensitive to the filling ratio. The high-frequency MSB, 
which is an anti-crossing between two dispersion relations 
(fundamental mode-higher mode) will change its position according to the 
higher mode, which encounters the fundamental mode at higher frequencies 
as the filling ratio is increased. In other words, the fundamental 
mode is guided by the high index path, as for a dielectric waveguide, 
and the PBG is not important. On the contrary, the higher guided mode 
does exist because the field confinement is provided by the PBG.\\
We have undertaken dispersion relation 
calculations for values of $f$ where more than two MSBs exist. 
The higher-frequency MSB-1, see Fig. 3, is a band splitting of the 
fundamental mode at $k=\pi/a$, whereas the higher-frequency MSB-2 
is not a neat MSB, but a more complicated coupling among the 
guided modes and the photonic bands at the bulk gap's edge.
Moreover, from transmission studies, we have noticed a similar 
behaviour for the first and third MSBs, and for the second and fourth 
MSBs respectively. 

We have observed that the transmission through single-defect PC 
waveguides has very small values in narrow frequency regions, 
the so-called MSBs. There are two types of MSBs and they are due 
to the 1D periodic potential of the PC along the waveguide. 
One MSB comes from the splitting of the folded 
fundamental guided mode at the zone edge of the reduced Brillouin zone; 
the other one comes from the anti-crossing between the fundamental mode 
and the higher order mode. The fundamental mode is sustained by a 
index guiding mechanism, whereas the higher modes are distinctive of 
PC waveguides and they are confined by the PBG. This difference is found 
in the dependence of the MSBs' position on the filling ratio as well 
as in the dispersion relation. 

We would like to thank S.\ Foteinopoulou, L.C.\ Andreani and 
S.G.\ Johnson for helpful discussions. This work was supported by 
the IST project PCIC and NSF Grant No. INT-0001236.
Ames Laboratory is operated for the U.S.\ Department of Energy
by Iowa State University under Contract No. W-7405-Eng-82.


\begin{references}
\bibitem{comment1}
Because of the coupling efficiency of the incident and outgoing 
waves with the guided mode, the peak transmission could result
less than one. To improve the coupling, we use a source with a 
spatial profile close to that of the fundamental guided mode of the 
PC waveguide. The fact that the transmission is independent of the 
length of the waveguide confirms that ``losses'' are not intrinsic 
and have to be referred to the non-optimal coupling efficiency.
\end{references}
\end{document}